# Trilobite tridents: A model of lift, drag and stability for queue formation

Hugh A. Trenchard[1], Carlton E. Brett[2], Matjaž Perc[3,4,5,6,7]

[1]805 647 Michigan Street, Victoria, BC V8V 1S9, Canada; h.a.trenchard@gmail.com
[2]Geology Department, University of Cincinnati, Cincinnati, OH 45221-0013, USA
[3]Faculty of Natural Sciences and Mathematics, University of Maribor, Koroška cesta 160, 2000 Maribor, Slovenia
[4]Community Healthcare Center Dr. Adolf Drolc Maribor, Ulica talcev 9, 2000 Maribor, Slovenia
[5]Department of Physics, Kyung Hee University, 26 Kyungheedae-ro, Dongdaemun-gu, Seoul 02447, Republic of Korea
[6]Complexity Science Hub, Metternichgasse 8, 1030 Vienna, Austria
[7]University College, Korea University, 145 Anam-ro, Seongbuk-gu, Seoul 02841, Republic of Korea

## Abstract

The bizarre trident-like cephalic projections of *Walliserops trifurcatus* have previously been interpreted as sexually selected weapons for intraspecific combat. We evaluate one possible, non-exclusive alternative hypothesis grounded in biomechanics and collective behaviour: that tridents contributed to hydrodynamic lift and queue stability, thereby conferring energetic and mechanical advantages during group locomotion. Other possible functions, including defence, sensory use, tactile interaction, or combinations of these roles, are not tested directly in the present model. Under this hypothesis, lift could offset gravitational forces, enabling greater locomotor efficiency, while mechanically linked formations—where tridents rested on the pygidia of leading individuals—enhanced pitch and roll stability and minimized costly accelerations and collisions. These formations may also have facilitated hydrodynamic drafting, which preceded trident evolution, allowing weaker individuals to conserve energy and remain integrated within the group. With power output held constant, the additional drag associated with the elevated, lift-producing configurations outweighed the reduction in apparent-weight-related demand, leaving modeled speeds below the streamlined 0° no-lift baseline while increasing the proportion of power potentially saved through drafting. The trident's structure, though potentially costly for solitary locomotion or rapid combat movements, may have functioned effectively in cooperative formations, suggesting that its original selective advantage lay not in individual performance but in enhancing group cohesion and efficiency. This interpretation implies morphological specialization evolving from collective energy-saving behavior among conspecifics. Rather than emerging from competitive pressures alone, the trident may have arisen through selection for coordinated, cooperative movement—potentially representing a precursor stage to traits later exapted for sexual selection.

## Introduction

Self-organized queue formations are common in nature. Examples of such formations include inanimate multi-particle chains (Šiler et al., 2012), ants in two-way traffic (Wang et al., 2020), spiny lobsters on the sandy bottoms of the Great Bahama Bank (Bill and Herrnkind, 1976; Herreid, 2012), processionary caterpillars (Scoble, 1992), birds in V-formation (Lissaman and Shollenberger, 1970), competitive cyclists (Trenchard et al., 2014), people queueing at bus stops or service counters (Kneidl, 2016; Leibowitz, 1968) and even in non-physical ways such as animal social queues in which subordinates wait their turn to inherit dominant breeding status (Kokko and Johnstone, 1999; Wong et al., 2007).

In the present benthic context, a queue refers to an approximately linear, head-to-tail arrangement of trilobites progressing along the substratum; drafting refers to the reduction in hydrodynamic drag and power demand experienced by individuals occupying the wakes of those ahead; and positional stabilization refers to the maintenance of relative spacing, directional alignment, and pitch–roll orientation within the formation through coordinated movement and, where applicable, trident–pygidium contact.

Evidence of queue formations also appears in the fossil record. *Synophalos xynos* formed chain-like associations (Hou et al., 2009), as did *Waptia*-like arthropods (Hou et al., 2008). The trilobites *Ampyx priscus* Thoral, 1935 (Vannier et al., 2019) and *Bathycheilus castilianus* Hammann, 1983 (Gutiérrez-Marco et al., 2009), together with representatives of the genera *Agerina* (Chatterton and Fortey, 2008) and *Trimerocephalus* (Radwański et al., 2009; Błażejowski et al., 2016), are preserved in fossilized queue formations.

Recently, Gishlick and Fortey (2023) considered the evolutionary function of cephalic anterior tridents on the trilobite *Walliserops trifurcatus* Morzadec 2001, from the Lower Devonian (Upper Emsian, lower nodular beds in exposed section of the Timrhanrhart Formation (ZGEE1)), northeast of Jbel Gara el Zguilma, near Foum Zguid, southern Morocco (Chatterton et al., 2006) and proposed that *Walliserops* used their tridents for combat behavior. The authors argued that this behavior indicates sexual selection and attempted to rule out other possible functions such as defence, a sensory aid to feeding, or a stirring probe to agitate sediment and disturb prey items. They further argued that a teratological malformation (a fourth tine) in one specimen would have impaired defensive or sensory functions but not hindered combat effectiveness, supporting a sexual selection hypothesis.

The energy-saving hypothesis examined here is only one of several possible functional interpretations. Among other things, the trident could also have contributed to defence, sensory exploration, tactile interaction between conspecifics, or combinations of these roles. These possibilities are not necessarily mutually exclusive, and the trident may have served different or overlapping functions during its evolutionary history. The present study does not attempt to test all possible interpretations or exclude other functions; rather, it develops and evaluates one specific biomechanical model involving energy-saving by hydrodynamic drafting and close following, lift, and queue stability.

Within this broader range of possible functions, we hypothesize that the *Walliserops* trident may have acquired adaptive value through a combination of hydrodynamic lift, drafting and mechanical stabilization during group locomotion. Lift could reduce apparent-weight-related locomotor costs, while the drag associated with an elevated trident–body configuration could increase the energetic value of occupying a leader's wake. Trident–pygidium contact could also counteract instability and help followers maintain drag-reduced positions. The adaptive value of trident elongation may therefore have emerged primarily during collective locomotion.

Although lift may have been irregular due to turbulent fluid flow over and between tridents, collective stability among *Walliserops* may have been enhanced by resting their tridents on the pygidia of those ahead. In this way, tridents provided pitch and roll stability in queue formations, especially for individuals that occupied following positions during travel, but also for queue leaders.

Similarly, we propose that tridents aided positional modulation in queue formations by allowing individuals to maintain a roughly constant distance between each other during queue migration. This reduced energetically costly collisions and variations in speed and direction during collective propulsion. Further, at sufficient locomotion or water current speeds, queue formations served to reduce drag for following trilobites and to reduce associated energy costs.

We refer to four species of *Walliserops* that bore a distinctive three-tined anterior cephalic "trident," in contrast to the single, pointed median cephalic projection of *Ampyx*: *W. trifurcatus*, *W. hammii* and *W. tridens* Chatterton et al., 2006 (Upper Emsian, lower nodular beds in exposed section of Timrhanrhart Formation (ZGEE1)), northeast of Jbel Gara el Zguilma, near Foum Zguid, southern Morocco (Chatterton et al., 2006). The fourth is *W. lindoei* Chatterton and Gibb 2010, Middle Devonian (Eifelian) from the Erbenochile Bed, basal El Otfal Formation, Jbel Issomour, near Alnif, southeastern Morocco. Together, these occurrences place *Walliserops* within the broader Anti-Atlas region, which encompassed diverse marine environments ranging from shelf and platform settings to deeper basins (Bault et al., 2022; Wendt, 2021). The present analysis, however, focuses on *W. trifurcatus* and uses measurements based on a fossil specimen illustrated by Gishlick and Fortey (2023).

Taxonomically, these four *Walliserops* species can be divided into two morphological subsets: *W. trifurcatus* and *W. lindoei*, whose tridents possess distinct shafts, leaflike tines, and midribs; and *W. hammii* and *W. tridens*, which bear tridents lacking shafts and exhibiting narrower tines without midribs. These structural differences raise the possibility that the two subsets may ultimately represent different genera and may also indicate functional divergence in trident use. A full taxonomic reassessment, however, lies beyond the scope of this paper. Accordingly, for purposes of the present analysis, *W. trifurcatus*, *W. lindoei*, *W. hammii*, and *W. tridens* are all treated as *Walliserops* species**.**

Based on eye morphology and hypothesized light perception capacity, *Walliserops* are thought to have dwelt at moderate ocean depths, near but above the base of the photic zone, below the storm wave base for the region (Chatterton et al., 2006). The photic zone extends as deep as 130–140 m (Gundersen et al., 1976), where ocean currents can average 30 cm $s^{-1}$ (Kutsuwada and Inaba, 1995; Ueki et al., 1998) to 35 cm $s^{-1}$ (Huyer et al., 1997) or greater (Kutsuwada and McPhaden, 2002).

During locomotion *Walliserops* likely encountered varying ocean current speeds and topography. These changing environments imply corresponding variation in *Walliserops* traveling speeds and metabolic output. Facing into a current would have increased the metabolic cost of forward motion at any given speed relative to still water, whereas moving with a rearward current would have reduced the metabolic cost. Similarly, advancing up slopes was metabolically more costly than traveling down slopes.

## Materials and Methods

### Comparative locomotor performance and model speeds

When discussing the energy expenditure and self-propulsion dynamics of any organism, it is useful to understand its metabolic demands and scope, specifically in terms of power output. This is difficult to determine for extinct organisms, but comparisons with living analogs permit reasonable estimates. We use these comparisons to select biologically plausible speeds corresponding to moderate- and high-output model conditions, rather than to reconstruct the species' actual locomotor performance.

Body mass is generally positively related to absolute metabolic capacity, including maximum metabolic rate, although the relationship is often less than directly proportional and varies with physiology, lifestyle and locomotor mode (Killen et al., 2007; Rubio-Gracia et al., 2020). Among aquatic animals, body mass can also predict swimming capacity within species, but speed additionally depends on body shape, muscle proportion, propulsive mechanics, effective reference area and hydrodynamic drag (Rubio-Gracia et al., 2020). Consequently, the greater estimated mass of *W. trifurcatus* relative to *Trimerocephalus chopini*, as discussed in Trenchard et al. (2017), Song et al. (2021), and Wang et al. (2024) suggests greater absolute power capacity, but does not by itself demonstrate a proportionately or substantially greater locomotor speed. We therefore use comparisons with similarly sized benthic crustaceans to define plausible speed scenarios rather than deriving *W. trifurcatus* speed directly from body mass.

We first estimated the mass of our analyzed hypothetical *W. trifurcatus* specimen by approximating its body as an ellipsoid. Using length ($L$) = 4.3 cm, width ($W$) = 2.7 cm, height ($H$) = 1.0 cm (Gishlick and Fortey, 2023, Fig. 3), we find its volume ($Vol$):

$$Vol = \frac{4}{3}\pi \times \left(\frac{L}{2}\right) \times \left(\frac{W}{2}\right) \times \left(\frac{H}{2}\right) . \quad (1)$$

This yields $Vol = 6.079 \times 10^{-6}\ \mathrm{m}^3 (6.079\ \mathrm{cm}^3)$. Mass was estimated as $m = \rho_{\mathrm{trilobite}} \times Vol,$ where $m$ is mass and $\rho_{\mathrm{trilobite}}$ is the assumed bulk density. Using $\rho_{\mathrm{trilobite}} = 1140\ \mathrm{kg\,m^{-3}}$ , adopted from the *T. chopini* model of Wang et al. (2024), gives $m = (1140\ \mathrm{kg\,m^{-3}})(6.079 \times 10^{-6}\ \mathrm{m}^3) = 0.00693\ \mathrm{kg} = 6.93\ \mathrm{g}$.

The actual bulk density of *W. trifurcatus* is unknown and may have differed because trilobite taxa varied in exoskeletal thickness, mineralization, and body composition. Accordingly, 6.93 g is treated as a reasonable estimate conditional on the assumed density.

Due to seawater buoyancy, we estimated the apparent weight of the submerged animal by accounting for density differences between trilobite composition and seawater:

$$G' = (\rho_{\mathrm{trilobite}} - \rho_{\mathrm{fluid}})\, g \times Vol, \quad (2)$$

where $G'$is apparent weight, $\rho_{\mathrm{trilobite}}$ is the assumed bulk density of the trilobite, $\rho_{\mathrm{fluid}} = 1025\ \mathrm{kg\,m^{-3}}$ is seawater density, $g = 9.81\ \mathrm{m\,s^{-2}}$ is gravitational acceleration, and $Vol$ is displaced volume. This yields an apparent weight of $G' = 0.00686$ newtons (N), or approximately 0.0069 N.

Extant crustaceans spanning the estimated body mass of *W. trifurcatus* provide useful benchmarks for selecting model speeds (Table 1), although differences in morphology, gait and locomotor medium preclude direct transfer. Ghost crabs *Ocypode guadichaudii* (2.78 ± 0.61 g) attained steady-state aerobic metabolism while running at 3.61–7.78 cm $s^{-1}$ during treadmill trials lasting up to 20 min (Full and Herreid, 1983). Larger adult crayfish *Procambarus clarkii* (22.8 ± 3.3 g; approximately 40-mm carapace length) walked underwater at a mean speed of 18.54 ± 2.81 cm $s^{-1}$ during spontaneous aquarium movements rather than controlled endurance trials (Drozdz et al., 2006). The estimated 6.93 g mass of *W. trifurcatus* lies between these two comparisons, whose observed speeds bracket the adopted moderate-speed value. They therefore support approximately 10–20 cm $s^{-1}$ as a plausible range for submerged, multi-limbed benthic locomotion, but not as a formal mass-based prediction.

Swimming studies provide broader performance bounds. Whiteleg shrimp *Litopenaeus vannamei* weighing 3.6–4.9 g were tested against currents of 5.41–11.47 cm $s^{-1}$, with endurance declining as speed increased (Zhang et al., 2006). Kuruma shrimp *Marsupenaeus japonicus* weighing approximately 10–11 g had a critical swimming speed of 32.87 ± 0.53 cm $s^{-1}$ and were tested at 23.0–38.6 cm $s^{-1}$ (Yu et al., 2009a, 2009b). Mean endurance declined from approximately 72 min at 23.0 cm $s^{-1}$ to approximately 18 min at 38.6 cm $s^{-1}$, supporting a general distinction between longer-duration movement at lower speeds and shorter-duration movement near 40 cm $s^{-1}$.

We therefore use rounded values of 11 and 40 cm s⁻¹ as representative moderate- and high-output model speeds, approximately corresponding to walking and rapid locomotion, respectively. The lower value lies between the speeds observed in the two benthic comparisons and is close to the highest flow speed tested for lower-mass *L. vannamei*. The higher value approximates the highest test speed for higher-mass *M. japonicus*. These comparisons bracket the estimated mass and proposed performance range of *W. trifurcatus* but do not establish a mass-based scaling relationship. The adopted values are illustrative inputs for evaluating the velocity-dependent hydrodynamic model and should not be interpreted as known sustainable or maximal speeds of *W. trifurcatus*.

**Table 1**. Locomotor performance of extant crustaceans used to establish biologically plausible model-speed bounds for *Walliserops trifurcatus.*

| Taxon, mass/size and reference | Locomotor mode and setting | Reported performance | Comparative relevance and limitations |
|---|---|---|---|
| *Ocypode guadichaudii* (painted ghost crab) **Mass/size:** 2.78 ± 0.61 g **Reference:** Full and Herreid (1983) | Terrestrial, lateral multi-legged running in a treadmill respirometer. | Aerobic steady-state exercise at 0.13, 0.19, and 0.28 km h⁻¹ (3.61, 5.28, and 7.78 cm s⁻¹) during 10- and 20-min trials. Speeds below 0.30 km h⁻¹ (8.33 cm s⁻¹) were sustained for more than 40 min. | Provides a relatively close mass comparison and evidence that a small decapod can sustain multi-legged locomotion at several centimetres per second. Because locomotion was terrestrial, the results do not incorporate submerged hydrodynamic costs. |
| *Procambarus clarkii* (red swamp crayfish) **Mass/size:** Adults; approximately 4 cm carapace length, 22.8 ± 3.3 g **Reference:** Drozdz et al. (2006) | Submerged walking and exploratory behavior in a freshwater aquarium during 20-min observations. | Mean walking speed of 18.54 ± 2.81 cm s⁻¹ ; 18.06 ± 4.47 cm s⁻¹ under two different experimental conditions. | Demonstrates that submerged, multi-legged benthic locomotion can occur at speeds on the order of 10–20 cm s⁻¹. The reported value was an average behavioral speed rather than an endurance or fatigue measurement, and body mass was unavailable for direct comparison. |
| *Litopenaeus vannamei* (whiteleg shrimp) **Mass/size:** 3.6–4.9 g wet mass; 8.3–8.8 cm total length **Reference:** Zhang et al. (2006) | Pleopod-driven swimming against current in a flume. | Tested at 5.41, 6.78, 8.21, 10.11, and 11.47 cm s⁻¹ for up to 9000 s; endurance decreased with increasing speed. | Provides a close mass comparison and direct aquatic evidence supporting approximately 11 cm s⁻¹ as a plausible moderate-flow model condition. Pleopod-driven swimming differs mechanically from benthic trilobite locomotion. |

| Taxon, mass/size and reference | Locomotor mode and setting | Reported performance | Comparative relevance and limitations |
|---|---|---|---|
| *Marsupenaeus japonicus* (kuruma shrimp) **Mass/size:** 10.18 ± 1.89 g wet mass; 9.92 ± 0.60 cm body length **Reference:** Yu et al. (2009a) | Pleopod-driven swimming, predominantly near the bottom of a flume. | Critical swimming speed, Ucrit, of 32.87 ± 0.53 cm $s^{-1}$; swimming examined at 23.0–38.6 cm $s^{-1}$. | Provides a broadly comparable mass range and demonstrates high-output aquatic locomotion at approximately 30–40 cm $s^{-1}$ in an extant crustacean. It does not establish equivalent performance in a benthic trilobite. |
| *Marsupenaeus japonicus* (kuruma shrimp) **Mass/size:** 11.04 ± 2.43 g wet mass **Reference:** Yu et al. (2009b) | Pleopod-driven swimming against current in a flume until fatigue. | Mean endurance declined from 4340 s (approximately 72 min) at 23.0 cm $s^{-1}$ to 1067 s (approximately 17.8 min) at 38.6 cm $s^{-1}$. | Supports approximately 40 cm $s^{-1}$ as a conditional, relatively short-duration high-output model scenario. It does not support a reconstructed maximum speed of *W. trifurcatus*. |

## Forces and moments acting on *Walliserops*

We consider four primary forces: gravity, hydrodynamic drag, lift and buoyancy. Drag acts parallel but opposite to the direction of travel. Buoyancy reduces gravitational loading to the apparent weight defined by Eq. (2), whereas upward lift further reduces the residual vertical load and downward lift increases it. Buoyancy is determined by the volume, shape and density of an organism. Lift arises from a pressure differential between the ventral and dorsal surfaces, consistent with Bernoulli's principle, together with the downward deflection and acceleration of seawater by the body and trident, which produces an equal and opposite upward reaction force in accordance with Newton's third law. Its magnitude depends primarily on body and trident shape, exposed surface area, angle of attack, relative flow velocity, fluid density, and whether the surrounding flow is laminar, transitional, or turbulent (Hoerner, 1965). Moments describe the tendency of forces to induce rotational motion in the organism. These forces and moments are summarized in Figure 1, in relation to *W. trifurcatus*.

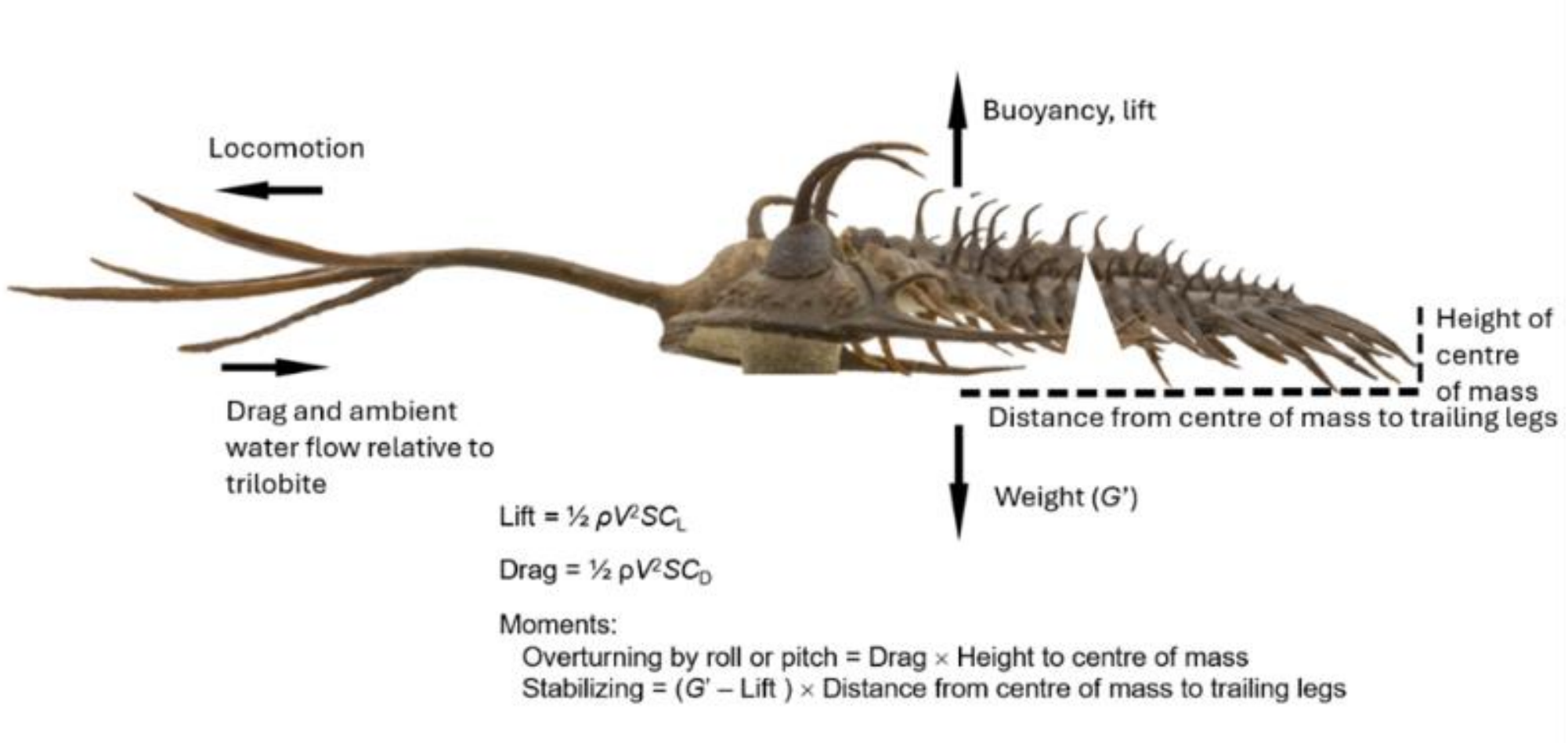

**Figure 1.** Forces and moments acting on *W. trifurcatus* during locomotion. Roll refers to rotation about the anterior-posterior axis, while pitch refers to rotation along the medial-lateral axis. *W. trifurcatus* image adapted from Gishlick and Fortey (2023), Fig. 1 (Creative Commons licence). Open notch indicates hypothetical straightening of a partially enrolled specimen. Figure inspired by Martinez et al. (1998, fig. 2).

## Drag forces

Drag is the moving fluid resistance as it contacts an object, described by:

$$F_D = \frac{1}{2}\left(\rho \times V^2 \times S \times C_D\right) \tag{3}$$

where $F_D$ is the drag force, $\rho$ (rho) is the fluid density (seawater 1025 kg/m$^3$); V is the velocity of the object relative to the fluid; $S$ is the cross-sectional area of the object facing the flow (here 2.12 cm$^2$); $C_D$ is the drag coefficient.

Total drag would not, however, be determined by frontal area alone. The spiny dorsal exoskeleton, trident, body curvature, surface roughness, appendages, and orientation relative to the flow would alter pressure distribution, flow separation, wake formation, and skin-friction drag. In the conventional drag equation, many of these shape-dependent effects are represented collectively by $C_D$, although their individual contributions cannot be resolved without a three-dimensional reconstruction and species-specific computational or experimental analysis.

Using computational fluid dynamics, Song et al. (2021) obtained drag coefficients for lead-position *T. chopini,* a smaller trilobite than *W. trifurcatus*, that varied between dimensionless 0.91 and 1.45 (mean 1.11) depending on their speed. We use Hoerner's (1965) drag coefficient for cylinders of 1.17, which is close to Song et al.'s (2021) $C_D$ mean. The model therefore treats the estimated frontal body area and an assumed bulk drag coefficient as approximations. The baseline 0° calculation does not separately include projected trident or dorsal-spine area or their associated flow disturbance. The trident contribution to the effective area of elevated configurations is introduced later through $A_{eff}$ in Eqs. (10)–(11). More precise estimates would require CFD or physical-model testing based on a complete three-dimensional reconstruction of *W. trifurcatus*.

Power output is more useful than speed when considering the different forces that affect self-propulsion speeds and the hydrodynamic consequences of those speeds. The model combines the power required to overcome hydrodynamic drag with a simplified apparent-weight-related power proxy. The latter represents otherwise unquantified load-dependent costs associated with body and limb load-bearing, body orientation, substrate interaction and the associated muscular and metabolic demand; it is not literal mechanical work performed against gravity on a level substratum or a complete metabolic-energy budget. At the illustrative locomotion speed ($LS$) of 11 cm s$^{-1}$, combined modeled power is therefore expressed as:

$$P_{\text{total}} = P_{\text{apparent}} + P_{\text{drag}} \tag{4}$$

where $P_{\text{apparent}} = G' \times LS$, and $P_{\text{drag}} = F_D \times LS$.

Proportionality between the apparent-weight-related costs and $G' \times LS$ is an explicit simplifying assumption of the model.

Substituting $F_D$ from Eq. (3) and taking $V = LS$, where $V$ is the relative fluid flow speed, this can be written as:

$$P_{\text{drag}} = \frac{1}{2} \rho\, V^3\, S\, C_D. \tag{5}$$

Following Wang et al.'s (2024) application of the submerged Froude number ($Fr$) , we use $Fr \approx 1$ as an operational threshold identifying an approximate transition into a regime in which hydrodynamic drag becomes sufficiently important for drafting to provide a meaningful energetic benefit. This is not a sharp physical cutoff: below the transition, drafting may still reduce drag, but the saving represents a comparatively limited proportion of total modeled power demand; above it, drag accounts for an increasing share of power demand and the potential energetic benefit of drafting increases, as evaluated in the "Hydrodynamic drafting" and "Model results" subsections. In the present model, $Fr = 1$ occurs at approximately 20.63 cm s$^{-1}$.

## Locomotion in currents

Positive rheotaxis occurs when an organism orients and progresses against the current. Let $V_o$ denote the organism's forward speed relative to the substratum and $V_c$the magnitude of the opposing current. The relative flow speed experienced by the organism is then the sum of these velocities: $V_{\text{rel}} = V_o + V_c$.

Accordingly, the drag force during positive rheotaxis, $D_{\text{positive}}$, is:

$$D_{\text{positive}} = \frac{1}{2}\rho(V_o + V_c)^2 S C_D \tag{6}$$

Negative rheotaxis, or downstream progression, occurs when the organism moves in the same direction as the current. The relative flow speed is then the absolute difference between the organism's speed and the current speed: $V_{\text{rel}} = | V_o - V_c |$.

The corresponding drag force, $D_{\text{negative}}$, is:

$$D_{\text{negative}} = \frac{1}{2}\rho\, | V_o - V_c |^2\, S C_D. \tag{7}$$

Ocean currents therefore alter the relative flow speed, drag, and power required for an organism to maintain a given speed relative to the substratum. During positive rheotaxis, an opposing current increases relative flow speed and consequently increases the power required to maintain forward progression. During negative rheotaxis, a current moving in the same direction generally reduces relative flow speed and power demand, provided that the current speed does not greatly exceed the organism's own forward speed.

These factors provide context for the still-water power-output estimates evaluated in Figure 2. The 0.00094 W estimate at 11 cm $s^{-1}$ represents moderate walking on a level surface in still water. In still water the power required to overcome drag increases with the cube of relative flow speed. Maintaining the same speed against a head-on current would require substantially more power. A following current would generally reduce this requirement. Thus, a sufficiently strong opposing current could shift an otherwise moderate walking condition into a higher-drag regime, making drafting energetically consequential even though the animal's speed relative to the substratum remained unchanged.

## Hydrodynamic lift

Lift acts perpendicular to the direction of fluid flow. Upward (positive) lift adds to buoyancy and reduces the organism's effective weight, while downward (negative) lift increases its effective weight by pushing it toward the substratum (Martinez, 1996).

If the organism accelerates too rapidly, or if the angle of attack becomes too steep relative to its speed, it may stall—a condition marked by a sharp increase in drag and a sudden loss of lift, which can result in flipping (Davis et al., 2019). Stall typically occurs in hydrofoils between 15° and 20° (Shi et al., 2020; Alberti et al., 2023). Cetacean flukes are observed to stall at angles of attack > 20° (Fish et al., 2008), while angles of attack > 10° may cause horseshoe crab *Limulus polyphemus* to flip backward (Davis et al., 2019).

Trilobite lift dynamics have been investigated through computational fluid dynamics (CFD). For example, Pates and Drage (2024) showed that trinucleimorph trilobites used their genal prolongations to generate negative lift and resist overturning. Shiino et al. (2012, 2014) used CFD to propose that a forked ventral hypostome of nektobenthic *Hypodicranotus striatulus* Walcott 1875 assisted lift and helped stabilize swimming dynamics.

Lift is calculated using an equation analogous to drag, but with the lift coefficient $C_L$ replacing the drag coefficient:

$$L = \frac{1}{2}\rho \times V^2 \times S \times C_{\mathrm{L}} \tag{8}$$

The lift coefficient, $C_{\mathrm{L}}$, depends on body geometry, angle of attack, Reynolds number and the reference-area convention used to define the coefficient. Reynolds number characterizes the relative importance of inertial and viscous effects in a flow (Gaylord, 2007). The lift coefficient depends strongly on orientation and may be greatest at an intermediate angle of attack, becoming smaller at both lower and steeper angles, particularly following flow separation or stall (Gaylord, 2007; Anderson, 2011).

The corresponding relationship is unknown for *W. trifurcatus*. Blake (1985) reported a comparatively low lift coefficient of approximately 0.20 for the benthic crab *Lopholithodes mandtii*. We therefore adopt $C_{\mathrm{L}} = 0.20$ as a constant effective coefficient for plausible analysis rather than as a reconstructed coefficient for *Walliserops*. The model does not incorporate Reynolds-number or stall-dependent variation in $C_{\mathrm{L}}$.

When an organism in water generates a lift force equal to its submerged weight, or apparent weight (G') as defined in Eq. (2), the balance of vertical forces is zero (Blake, 1985):

$$G' - \frac{1}{2}\rho \times V^2 \times S \times C_L = 0 \quad (9)$$

where $\rho$ is the density of the surrounding fluid (seawater, 1025 kg/m$^3$); $V$ is the velocity of the object relative to the fluid; $S$ is the organism's cross-sectional area facing fluid flow. Therefore, at sufficient angles of attack and propulsion speeds, lift can fully counteract gravitational force, resulting in vertical force balance.

In the model, lift does not alter apparent weight ($G'$) itself; rather, it reduces the residual vertical load from $G'$to $G' - L$. When lift equals apparent weight ($L = G'$; Eq. 9), the residual vertical load—and therefore the modeled apparent-weight-related power component—reaches zero.

Thus, for a given elevated configuration—with angle of attack, effective area and hydrodynamic coefficients held constant—as lift reduces the net apparent-weight-related load ($G' - L$) toward zero, less power is required for weight-related locomotor costs. At fixed total power, more power could therefore be directed toward overcoming horizontal drag, potentially enabling a higher speed than the same elevated configuration could attain without lift. This does not imply a higher speed than in the streamlined 0° condition, because the additional drag of the elevated configuration may outweigh the benefit of lift.

A comparable lift-based explanation was proposed for the desert beetle *Onymacris plana*. Its flattened, laterally expanded body was hypothesized to act as an aerofoil because oxygen consumption increased with running speed up to 13 cm s$^{-1}$ but showed no significant further increase between 13 and 22 cm s$^{-1}$ (Bartholomew et al., 1985).

As a trade-off to the reduced effects of gravity, lift introduces greater surface area into the fluid stream, generating increases in drag. We determined the change in surface area according to angle of attack by:

(10)

$$A_{\text{eff}}(\theta) = A_{\text{frontal}} \cos(\theta) + A_{\text{ventral}} \sin(\theta) \quad (10)$$

where $A_{eff}$ is the wetted exposed reference-area proxy as the trilobite lifts its trident and body, $A_{frontal}$ is the estimated frontal surface area of 2.12 cm$^2$.

Ellipsoid ventral surface area is estimated by Knud Thomsen's approximation (Numericana, n.d.):

$$S \approx 4\pi \left[ \frac{a^p b^p + a^p c^p + b^p c^p}{3} \right]^{\frac{1}{p}} \quad (11)$$

where *a, b*, and *c* are the semi-axes of the ellipsoid, and exponent $p$ is typically taken to be around 1.6075 (www.numericana.com, retrieved July 2024).

The curved ventral body surface area was approximated as one-half of the surface area (wetted area) of a triaxial ellipsoid using the Knud Thomsen equation (Eq. 11), with semi-axes $a = 2.15$ cm, $b = 1.35$ cm, and $c = 0.50$ cm. The calculated full surface area is 21.66 cm², giving a ventral area of 10.83 cm². Increasing this estimate by 17% to represent the projected underside of the trident gives an effective ventral reference area of $A_{\text{ventral}} = 12.67$ cm². Using Eq. (10), illustrative angles of attack of 10°, 15°, and 20° give effective reference areas of 4.29, 5.33, and 6.33 cm², respectively; these values were used to calculate the power-equivalent speeds shown in Figure 3. A wetted-surface-derived effective area was used as a simplified proxy for surface-distributed hydrodynamic effects, including skin friction, that are not represented geometrically by projected planform area alone (Alexander, 1990; Gaylord, 2007).

All power-equivalent speeds reported in Figures 2–4 were calculated from the component relationships defined above. For the lift-adjusted calculations, the apparent-weight-related component was recalculated using the residual vertical load $G' - L$, while the drag component was recalculated using $A_{\text{eff}}$. Total modeled power was obtained by summing the resulting apparent-weight-related and drag-power components. Speeds were calculated algebraically where the applicable power relationship could be rearranged directly for speed; otherwise, the speed solutions were obtained numerically using Brent's bracketing method as implemented in SciPy (Virtanen et al., 2020). The resulting power–speed relationships and force-balance conditions are reported in Figures 2–4.

## Hydrodynamic drafting

In earlier work, Trenchard et al. (2017) proposed that queueing behavior among trilobites conferred energy savings during locomotion and migratory phases, allowing groups of these trilobites to travel farther and/or faster than if migrations occurred in the absence of this energy-saving mechanism (also Song et al., 2021).

Using CFD, Song et al. (2021) showed that energy savings from drafting diminish rapidly with increasing distance, becoming maximal when trilobites follow at near-contact distances. Accordingly, trilobites may have adopted near-contact following positions to maximize energy savings, aligning closely behind those ahead. Wang et al. (2024) found that *T. chopini* were too small and slow to generate substantial drag forces and thus incurred minimal energy costs during locomotion. However, this is not the case for *W. trifurcatus*, as evaluated in Figure 2.

Following Olds (1998), the drafting coefficient ($D_c$) is defined here as the ratio of the drag-related power required by a follower in a drafting position to that required by a non-drafting leader moving at the same relative flow speed. At the same speed, this is equivalent to the ratio of their drag coefficients ($C_{\text{D}}$). Trenchard et al. (2017) estimated $D_c$ using reported values for two cylinders in tandem: $C_{\text{D}} = 1.17$ for the leading cylinder (Hoerner, 1965) and $C_{\text{D}} = 0.45$ for the in-wake cylinder (Igarashi, 1981). The resulting ratio, $D_c = 0.385$, means that the follower is modeled as requiring 38.5% of the leader's drag-related power, corresponding to a 61.5% reduction in the drag-power component.

Applying $D_c$ to Eq. (3) gives the drafted drag force:

$$F_{\text{d, draft}} = \frac{1}{2}\left(D_c \times \rho \times V^2 \times S \times C_{\text{D}}\right) \qquad (12)$$

This estimated drag reduction aligns with empirical findings by Bill and Herrnkind (1976), who reported a 65% drag reduction in spiny lobsters moving in formation. It is also consistent with the 63% energy saving found by Fish (1995) for ducklings swimming behind a decoy compared to solitary swimming. These estimates do not include power required to overcome gravity or frictional forces, nor do they account for the effects of upslopes and downslopes on drag, speed and power.

Trenchard et al. (2017) treated the power required by the leader principally in terms of the drag-power component and did not explicitly model apparent-weight-related or other non-drag locomotor costs.

Accordingly, the estimated 61.5% saving applies only to the power required to overcome drag, not to the leader's complete locomotor-power requirement. The present model expands that treatment by incorporating apparent-weight-related and other non-drag requirements and by using $E_s$ to represent additional savings associated with those components. The resulting drag-component and combined modeled power savings are reported in the "Model results" subsection.

**Model of selection process**

Trenchard et al. (2017) introduced the following equation to model these dynamics:

$$TCR = \frac{S_{\text{front}} \times D_c}{MSO_{\text{follow}}} \tag{13}$$

which is mathematically equivalent to:

$$TCR = \frac{S_{\text{front}} - [S_{\text{front}} \times (1 - D_c)]}{MSO_{\text{follow}}} \tag{14}$$

where *TCR* is the "trilobite convergence ratio"; $S_{\text{front}}$ represents the power required by the lead trilobite to maintain its speed in the high-drag position. This power is the sum of the power required to overcome drag and all other non-drag components including apparent weight and other frictional forces. $S_{\text{front}}$ determines the pace for those following behind. $D_c$ is the drafting coefficient, as discussed above. $MSO_{\text{follow}}$ denotes the maximal sustainable power output of trilobites occupying drafting positions. $(1 - D_c)$ is the corresponding fractional reduction in drag power, expressed as a percentage: $(1-D_c) \times 100$. Equations (13) and (14) therefore define a deliberately drag-based convergence ratio rather than a complete locomotor-power requirement. Equations (13)–(15) describe relative power capacity and group cohesion; they do not independently convert power into locomotor speed. For a specified trident configuration, angle of attack and set of environmental conditions, the speed corresponding to a given power output is obtained from the applicable nonlinear power-to-speed equation.

Equations (13) and (14) model the relative drag-related power outputs between leading and following (drafting) trilobites. This model can be applied to characterize different phases of collective behavior in trilobite groups. Equations (13) and (14) generalize to groups of more than two trilobites, with the mean *TCR* value representing the collective state at a given time.

Extending this model and adapting equations (13) and (14) to incorporate additional energy savings conferred by tridents—such as reduced friction via lift forces and stabilization against costly decelerations and collisions—the term $E_s$ is added:

$$TCR = \frac{S_{\text{front}} - [(S_{\text{front,drag}} \times (1 - D_C)) + E_S]}{MSO_{\text{follow}}} \tag{15}$$

$E_s$ is the additional quantity of power saved through reductions in non-drag locomotor requirements, including apparent-weight-related loading, substrate friction, positional corrections, costly decelerations and re-accelerations, and other stabilization-related costs. Unlike $D_c$, $E_s$is not a dimensionless coefficient; it is a power quantity expressed in watts.

$S_{\text{front,drag}}$ is the component of that requirement attributable specifically to overcoming hydrodynamic drag. The remaining component, $S_{\text{front}} - S_{\text{front,drag}}$, represents apparent-weight-related and other non-drag locomotor requirements.

In the original TCR model, $D_c$ marks the lower boundary of a threshold zone extending to $TCR = 1$, while $1 - D_c$ defines the width of that zone. Within this range, followers are modeled as able to maintain the group's pace while drafting but unable to sustain an exposed overtaking position (Trenchard, 2015). Under the expanded treatment, the corresponding threshold zone depends on both the drag-power saving represented by $1 - D_c$ and the additional non-drag power saving represented by $E_s$.

When $TCR > 1$, weaker individuals are unable to generate the power needed to maintain the leaders' speed—even with the benefit of drafting—and consequently drift backward. Unless the leading trilobites reduce their pace to permit reintegration, the gap will widen progressively, ultimately resulting in the weaker trilobites becoming permanently dropped from the group. As discussed, this process illustrates a core selection mechanism governed by relative stamina.

Equations (13) through (15) indicate that greater energy savings allow weaker trilobites to maintain the pace set by stronger individuals, thereby supporting a wider range of metabolic capacities within the group. Conversely, when energy savings are low, group cohesion can only be maintained if all individuals' maximal sustainable power outputs fall within a narrow range, leading to more physiologically homogeneous groups.

## Results and discussion

### Model results: lift, drag and drafting

Wang et al. (2024) demonstrated that the trilobite *T. chopini* would only begin to encounter difficulty due to drag at 42 cm $s^{-1}$, a much higher speed than *T. chopini* was hypothesized to be capable of achieving, while ocean current speeds of this magnitude were rare or unlikely. *W. trifurcatus* represents a different case, however, such that hydrodynamic drag would have become significant within the assumed model-speed range, particularly where animal motion and ambient currents produced sufficiently high relative flow speeds. In addition, *W. trifurcatus* was larger than *T. chopini*, both in terms of mass and its body frontal surface area of ~2.12 $cm^2$, which is about 7.8 times greater than the frontal surface area of *T. chopini* of ~0.27 $cm^2$ (Trenchard et al., 2017). The greater frontal area indicates that *W. trifurcatus* would have encountered considerably greater drag than *T. chopini*. The resulting power relationships and modeled $Fr \approx 1$ condition for *W. trifurcatus* are shown in Figure 2.

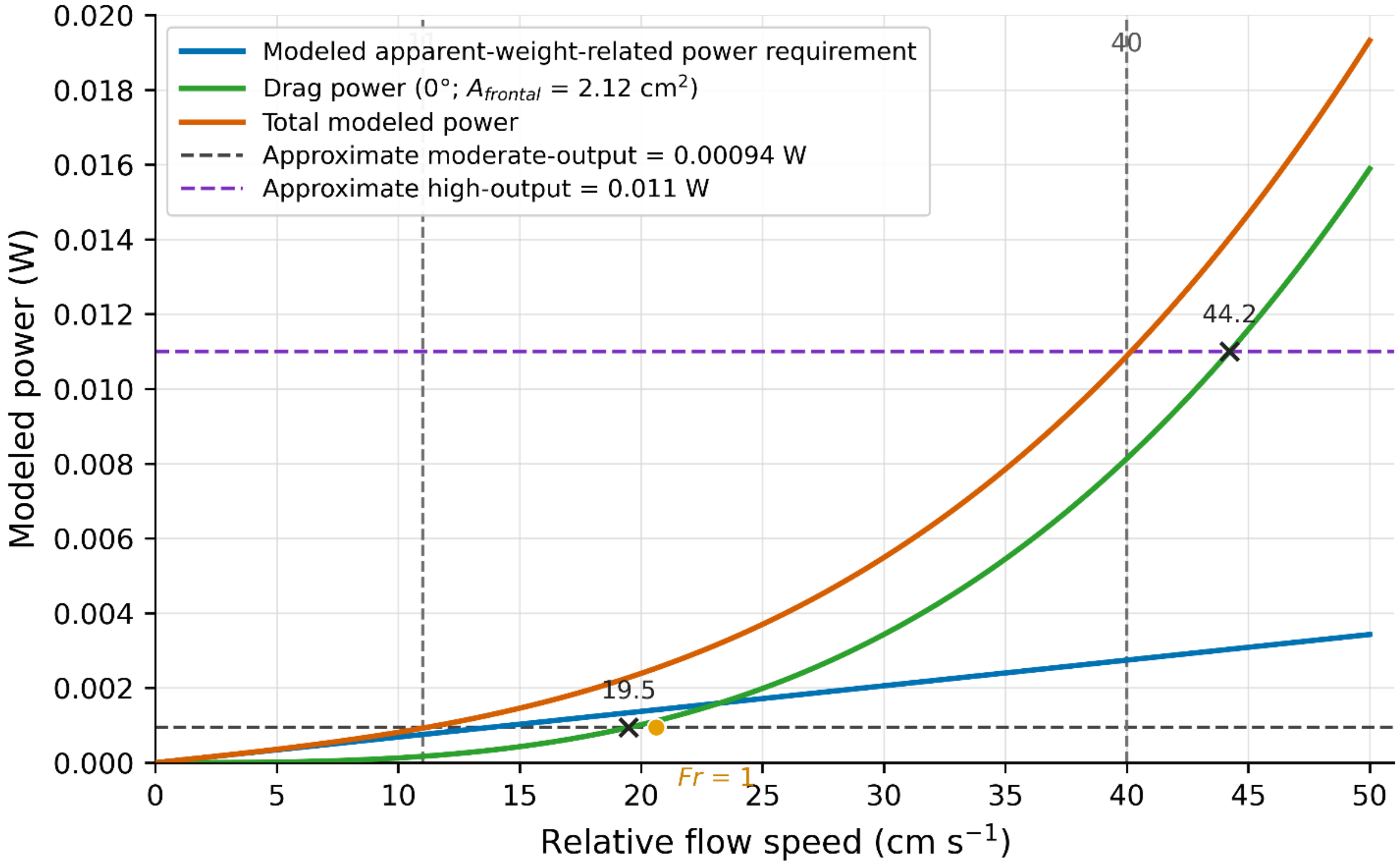

**Figure 2.** Modeled power requirements for *W. trifurcatus* as functions of relative flow speed in the streamlined 0° position. Curves show the apparent-weight-related requirement, the power required to overcome hydrodynamic drag, and their sum. Dashed lines and markers identify the approximate moderate- and high-output values and the modeled hydrodynamic transition conditions described in the text.

Figure 2 compares the modeled apparent-weight-related power requirement ($P_{\text{apparent}} = G \times {}'V$), the power required to overcome hydrodynamic drag ($P_{\text{drag}} = F_D \times V$), and their sum for the streamlined 0° no-lift condition. The drag calculation uses a frontal area of 2.12 cm² and $C_D = 1.17$. At relative flow speeds of 11 and 40 cm s$^{-1}$, combined modeled power is approximately 0.00094 and 0.011 W, respectively. The combined-power curve reaches these power outputs at approximately 11.14 and 40.18 cm s$^{-1}$. In a hypothetical drag-only condition—where apparent-weight-related and other non-drag costs are absent and the baseline $C_D$ and frontal area remain unchanged—the same outputs would support theoretical maximum speeds of approximately 19.48 and 44.23 cm s$^{-1}$.

Drag accounts for approximately 18.7% and 74.9% of combined modeled power at the respective moderate- and high-output intersections. The orange marker denotes the adopted $Fr \approx 1$ transition at 20.63 cm s$^{-1}$, where drag accounts for approximately 44.1% of combined modeled power. Because drag power increases with the cube of relative flow speed in still water, it becomes an increasingly dominant component of the modeled power requirement as speed rises. Consequently, the potential energy benefit of drafting, quantified later in this subsection, also increases with relative flow speed. These values are illustrative model outputs rather than measured physiological capacities.

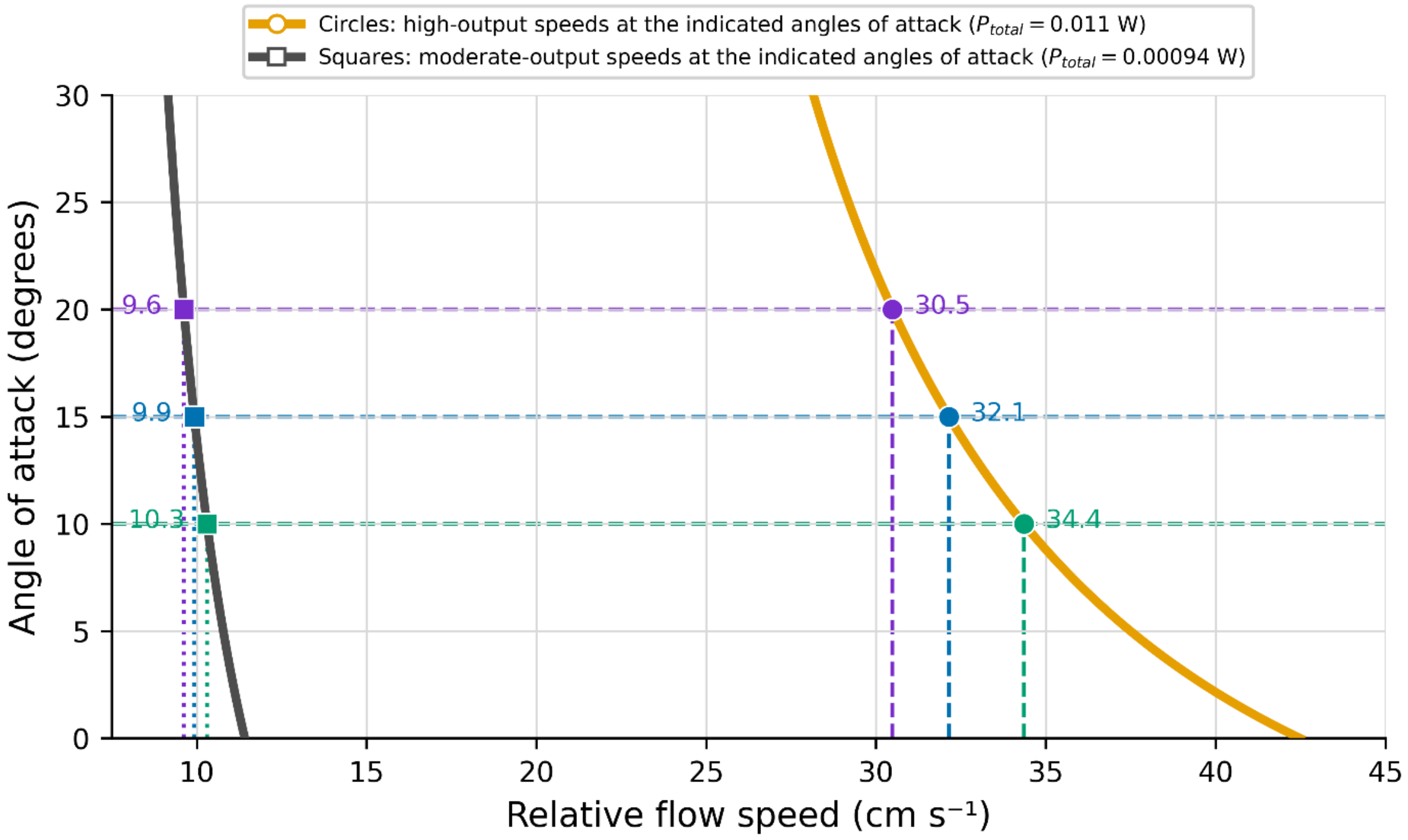

**Figure 3.** Modeled relationship between angle of attack and the relative flow speed at which lift-adjusted total power—comprising apparent-weight-related and hydrodynamic-drag components—reaches the modeled moderate-output (0.00094 W; gray curve) and high-output (0.011 W; orange curve) values. Squares and circles identify the respective moderate- and high-output intersection speeds at the illustrative 10°, 15°, and 20° angles of attack; adjacent values are speeds in cm $s^{-1}$.

At the moderate power output, the modeled speed decreased comparatively modestly from 10.3 cm $s^{-1}$ at 10° to 9.6 cm $s^{-1}$ at 20°. At the high power output, it decreased more substantially, from 34.4 to 30.5 cm $s^{-1}$. This difference reflects the strong dependence of both lift and drag on flow speed. At the moderate output, modeled lift was equivalent to only approximately 6.8–8.7% of apparent weight, whereas at the high output it was equivalent to approximately 75.6–87.8%. Lift therefore produced comparatively little reduction in apparent-weight-related loading at low speeds (power outputs) but became much more consequential at higher speeds.

The accompanying increase in drag nevertheless outweighed this reduction in apparent-weight-related loading in the fixed-power comparison. Drag represented approximately 29.9–35.9% of total modeled power at the moderate outputs but 94.8–97.7% at the high outputs. Consequently, increasing angle of attack reduced the attainable speed at a given total power despite the associated reduction in apparent-weight-related power demand.

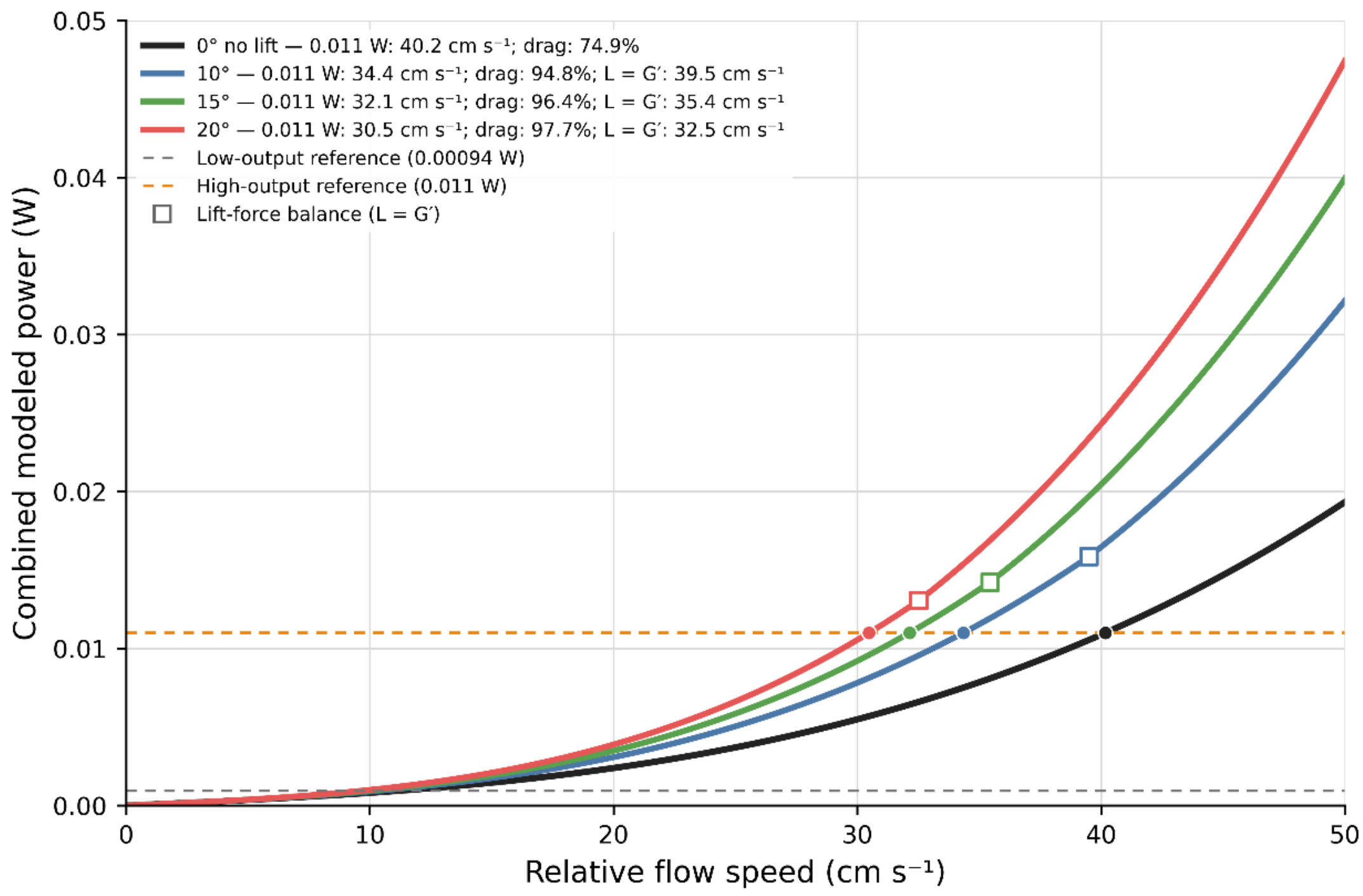

**Figure 4.** Combined modeled power versus relative flow speed for the streamlined no-lift baseline and the lift-adjusted 10°, 15° and 20° scenarios.

At the high-output intersections in Figure 4, the modeled speeds are 40.2 cm $s^{-1}$ for the streamlined no-lift baseline and 34.4, 32.1 and 30.5 cm $s^{-1}$ at 10°, 15° and 20°, respectively. Drag accounts for 74.9%, 94.8%, 96.4% and 97.7% of combined modeled power at these points. The open squares identify the vertical force-balance condition $L = G'$, which occurs at approximately 39.5, 35.4 and 32.5 cm $s^{-1}$ at 10°, 15° and 20°, respectively. Beyond these force-balance speeds, modeled lift exceeds apparent weight, and passive contact with the substratum can no longer be assumed.

Lift and drafting could therefore have provided complementary but distinct benefits during linked leader–follower locomotion. At each modeled trident–body angle of attack, including lift reduced the leader's apparent-weight-related power requirement and produced a higher modeled speed than the corresponding calculation at the same angle with lift excluded. At 10°, for example, lift increased the modeled speed at 0.011 W from 32.4 to 34.4 cm $s^{-1}$. This remained below the 40.2 cm $s^{-1}$ streamlined 0° no-lift result because the increased effective area of the elevated configuration imposed substantially greater drag. A closely following individual could nevertheless more readily maintain the leader's lift-assisted pace because drafting reduced the now-dominant drag component of its own power requirement.

This additional drag also increased the potential energetic value of drafting. Under the assumed drafting coefficient, a follower saves 61.5% of the power required to overcome drag, not 61.5% of the combined modeled power requirement. At the moderate output of 0.00094 W, this drag-power reduction corresponds to approximately 11.5% of combined modeled power in the streamlined 0° no-lift case and 18.4% in the 10° lift-adjusted case. At the high output of 0.011 W, the corresponding savings from combined modeled power are approximately 46.1% and 58.3%, respectively, as derived from the drag component at the corresponding modeled speeds shown in Figure 4.

The trident should therefore not be regarded as generating drafting itself; rather, an elevated trident–body configuration may have increased the energetic value of drafting while helping followers maintain close, stable wake positions. This trade-off supports a group-dependent, draft-facilitating interpretation of trident function, although the model does not establish that these combined benefits necessarily exceeded the additional drag cost.

In summary, Figures 2–4 illustrate complementary aspects of the modeled hydrodynamic trade-off, although the precise relationships cannot be reconstructed for *Walliserops*. Figure 2 establishes the streamlined 0° no-lift baseline and shows that hydrodynamic drag accounts for an increasing proportion of combined modeled power as relative flow speed rises. Figure 3 identifies the combinations of angle of attack and relative flow speed at which lift-adjusted power reaches the adopted moderate- and high-output values, while Figure 4 shows the corresponding power–speed relationships and vertical force-balance conditions. Applying the assumed drafting coefficient to the drag component further indicates that the percentage saving from combined modeled power increases as drag becomes a larger proportion of that requirement.

## Raised tines

The tridents of *W. trifurcatus* projected upward at an angle relative to the horizontal body plane, while *W. hammii*, *W. tridens* and *W. lindoei* lacked an elongate haft and did not show significant elevation (Gishlick and Fortey, 2023). *W. lindoei* tines, by contrast, appear only slightly inclined above the horizontal (Chatterton and Gibb, 2010), and all were likely capable of raising and lowering their cephala (Gishlick and Fortey, 2023), thereby allowing adjustment of tine angle.

The inclined tines and ventral body surface may have generated upward lift or some degree of asymmetric lift. Efficient and stable lift depends on the pressure distribution across a lifting surface (Anderson, 2011), and flow over and between the separated, rigid tines may have reduced lift efficiency or produced asymmetric and unsteady forces. In the model, lift reduced the residual vertical load and the associated apparent-weight-related power component. At sufficiently steep angles of attack, however, flow separation or stall could have produced an abrupt reduction in lift and increase in drag, while asymmetric forces could have promoted pitch or roll instability (Shi et al., 2020; Alberti et al., 2023; Fish et al., 2008; Davis et al., 2019). These effects may have imposed additional control and energy requirements on solitary individuals, particularly during forward movement or in variable currents.

Trident–pygidium contact within a queue could have mitigated some of this instability. By resting its trident against the posterior region of the individual ahead, a follower may have formed a mechanically connected queue in which contact transmitted reaction forces and limited relative pitch, roll and lateral displacement. More stable alignment could also have helped followers remain within reduced-drag wake positions while reducing collisions, directional corrections and repeated speed adjustments. Individuals whose tines improved such contact and stabilization might consequently have gained an advantage during collective locomotion, although the extent of torque transmission and stabilization remains to be tested experimentally.

These lift- and drag-related effects would have depended on relative water flow rather than movement over the substratum alone. Relatively stationary activities such as foraging, mating or moulting may therefore have involved lower hydrodynamic costs in weak or sheltered flows because the animals generated little relative flow through forward movement. Such activities would not, however, necessarily have been hydrodynamically unaffected: appreciable ambient currents could still have imposed lift, drag and destabilizing moments on an individual stationary relative to the substratum.

The morphology of *Walliserops* is compatible with the proposed following behavior. A follower's trident, held slightly above horizontal, could have rested against the pygidium or posterior thoracic region of the individual ahead. Because the body becomes lower and narrower toward the pygidium, the posterior exoskeleton of the individual ahead could have provided a shallow, tapering support surface for the follower's trident. This morphology may have helped keep the trident centred and limit lateral slippage without requiring the tines to extend as far forward as the cephalon. Figures 5 and 6 illustrate this proposed trident–posterior-body contact.

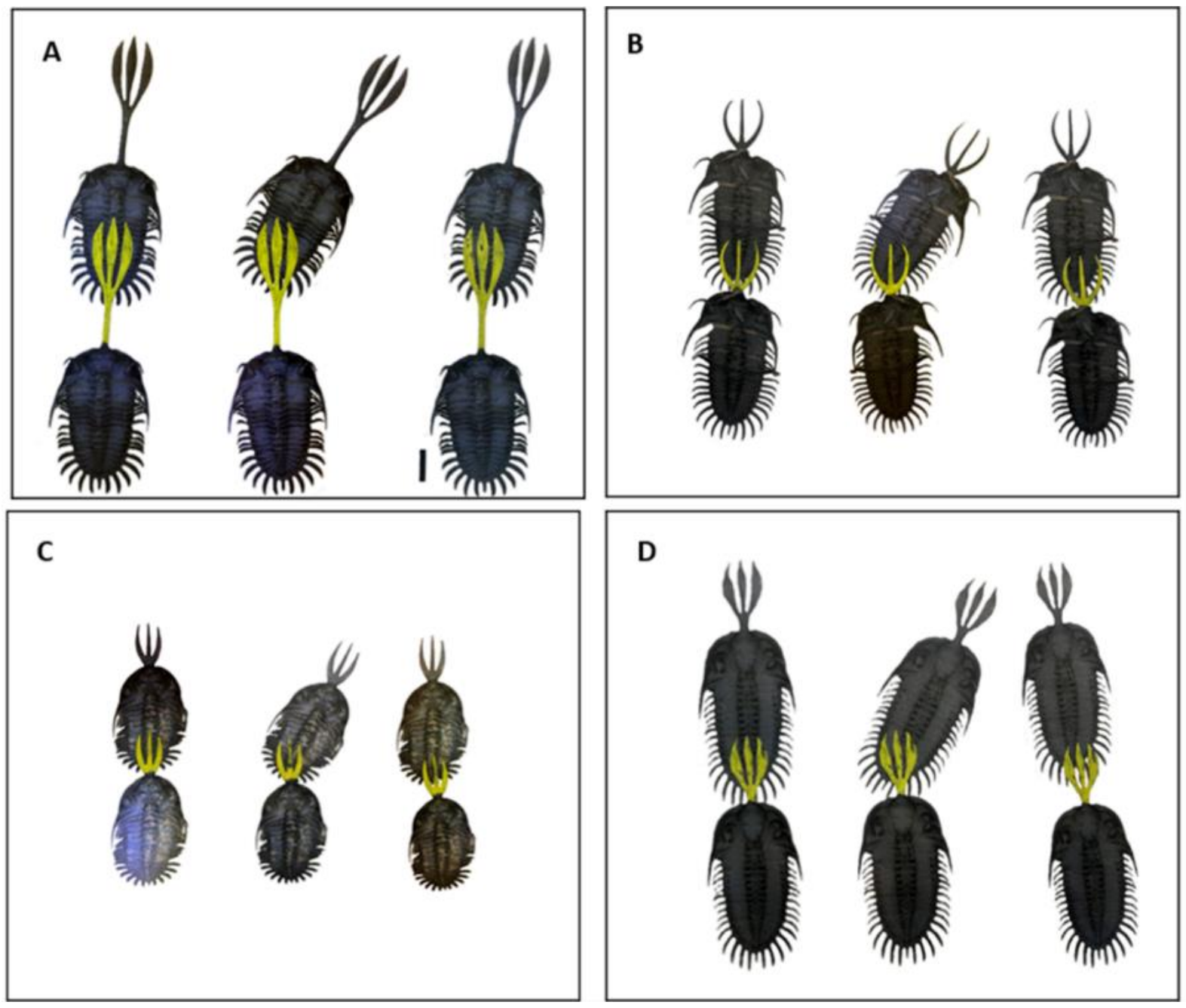

**Figure 5**. Four species of *Walliserops*: **A**. *W. trifurcatus* Morzadec 2001, UA 13447 (topotype) Lower Devonian (Upper Emsian); **B**. *W. hammii* Chatterton et al. 2006 (holotype) (Upper Emsian); **C**. *W. tridens*, Chatterton et al. 2006 (Upper Emsian) UA 13451 (holotype); **D**. *W. lindoei* Chatterton and Gibb 2010, Middle Devonian (Eifelian) ROMIP 56997. In each panel, the images model three hypothesized trident positions, dorsal view. Left: follower's trident rests on the leader's posterior, where the middle tine touches leader's axial lobe. Centre: leader rotated at approximately 30° relative to follower. Right: follower's trident rests on leader's posterior such that axial lobe is between two trident prongs. Color has been added to followers' tridents. Scale bar in A = 10 mm. Images adapted from fig. 1 of Gishlick and Fortey (2023) (Creative Commons licence).

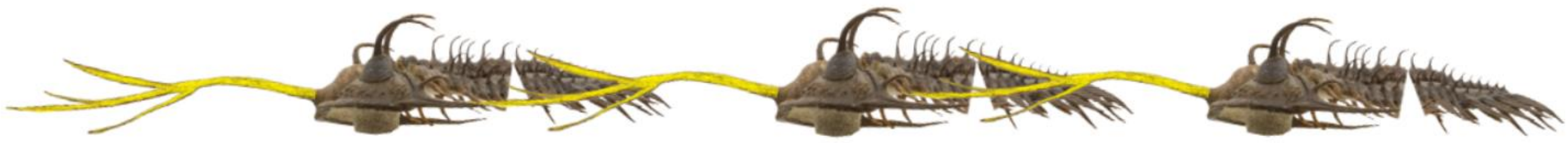

**Figure 6**. Model of *W. trifurcatus* in a linear following formation, lateral view. Image modified from Gishlick and Fortey (2023), Fig. S1 (Creative Commons licence).

Even without sensilla on the trident tines, as proposed by Gishlick and Fortey (2023), following trilobites could still have detected conspecifics ahead through mechanical strain detection. According to curved beam torsion theory (Young and Budynas, 2002), mechanical forces transmitted through exoskeletal contact would generate detectable strain at the trident base or cephalic joint (Zill et al., 2004; Wang et al., 2024). The teratological fourth tine observed in one *Walliserops* specimen (Gishlick and Fortey, 2023) would not have compromised this mechanosensory function.

We have not found published references to *Walliserops* in queue formations, which would support our hypothesis. However, Figure 7 shows an undescribed aligned cluster of four specimens of *W. lindoei*. Although they are not in contact, the two leftmost and the two rightmost *Walliserops* are aligned in opposite directions. This configuration is consistent with, though it does not demonstrate, the proposition that the follower's trident contacted the leader's posterior at times not quite captured in this fossil example. Even the slightest current disturbance during burial could have splayed them apart if they were dead; but it would be almost inevitable for the cephalic tines to rest on the posteriors of the ones in front in well-formed lines. Despite the weak evidence afforded by the specimen in Figure 7, it does show linear groupings and holds out the possibility for the discovery of better specimens.

Although we have not obtained specific information about the geographic origin of this specimen, it likely originates from similar Moroccan trilobite beds where *Walliserops* specimens have been found. The calcareous mudstone and muddy limestone in which these trilobites were preserved record a transitional setting between shallower, fully oxic and deeper lower energy and dysoxic environments. These beds yield a moderately diverse invertebrate community with small rugose corals, scattered small bivalves, gastropods, nautiloids, and very abundant trilobites, all with well-developed eyes (Chatterton et al., 2006 and Chatterton and Gibb, 2010; Brett et al., 2012a, 2012b).

This evidence indicates a normal salinity, relatively quiet, undisturbed offshore environment. Organisms lived on soft, muddy to slightly shelly substrates, well below wave base and below all but deepest storm wave effects. This habitat lay in the photic zone, as evidenced by well-developed eyes in phacopid trilobites. The substrate was soft, rather water-rich mud, based on burrow deformation. Benthic oxygenation probably ranged from lower oxic (~1 to 2 ml $L^{-1}$) to mildly dysoxic (~0.5 to 1 ml $L^{-1}$) (Brett et al., 2012a, 2012b).

Based on evidence for deep storm wave base and photic zone conditions, the fossilized trilobites in the specimen shown in Figure 7 were therefore likely living at depths of ~30 to 40 meters in a low-energy, soft, mud-bottom environment. Low-energy burial conditions were generally conducive to in situ or near-life position preservation. Based upon taphonomy, in some cases trilobites were caught up in bottom mud flows which obviously would have disrupted queues (Chatterton et al., 2006; Chatterton and Gibb, 2010; Brett et al., 2012a, 2012b). Conversely, in other cases the preservation of delicate moult ensembles (i.e., associations of exuviae of thoracopygidia and near proximity of cephalic shields) indicates rapid burial without any significant bottom disturbance (Type 1 obrution deposit; Brett et al., 2012a). Such conditions could potentially preserve trilobites in queued positions.

However, queues could only be preserved under very special conditions in which the cause of mortality involved sudden changes in conditions that caused near synchronous mortality; these trilobite chains literally dropped in their tracks then required subsequent rapid burial that was not accompanied by bottom disturbance, as in the case of moult ensembles. In slab CMC IP100892 shown in Figure 7, the presence of slight decay and incipient disarticulation in trilobite bodies suggests a short but still significant gap—days—between mortality and burial and very minor current movement to slightly disrupt the queue (Brett et al., 2012a; Brett et al., 2012b). However, it seems unlikely that the occurrence of four trilobites nearly in a row was produced by chance currents.

Notably, the individual at the far right of the slab in Figure 7 lacks a visible trident. Under Gishlick and Fortey's (2023) male-combat hypothesis, sexual dimorphism would be a reasonable expectation, with females potentially possessing smaller tridents or none at all. Gishlick and Fortey (2023) themselves noted that no definitive female morph has been identified and that tridentless or reduced-trident females might have been assigned to another taxon. However, the trident of the Figure 7 individual appears to have been severed during preparation (S. Rodgers, FossilEra.com, pers. comm., 2024), so this specimen cannot be used as evidence of a tridentless female. Moreover, the possibility that females also possessed tridents, perhaps differing from those of males only in size or form, cannot be excluded. The specimen therefore neither confirms nor refutes sexual dimorphism in *Walliserops*.

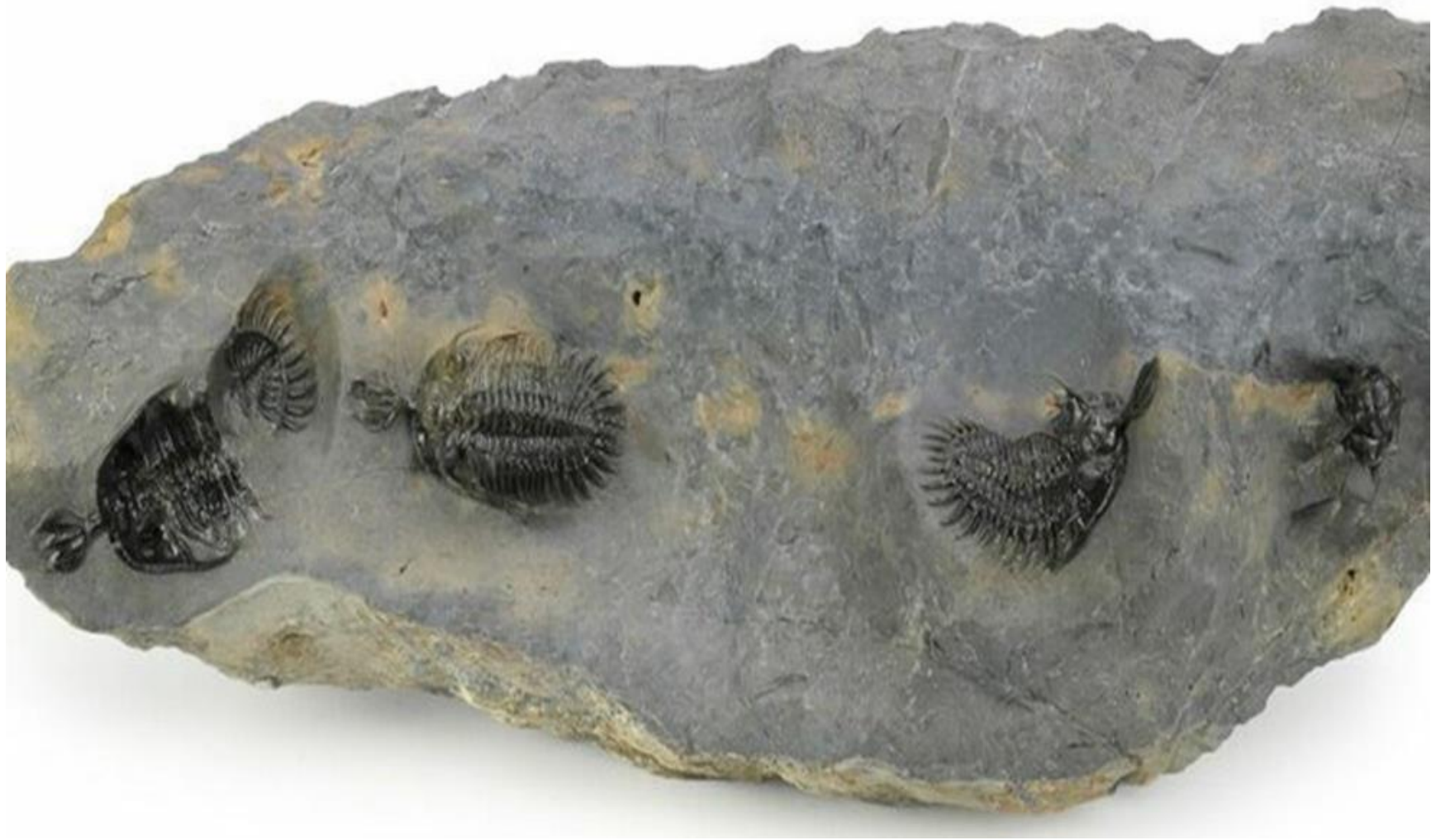

**Figure 7**. *Walliserops lindoei* Chatterton and Gibb 2010, Middle Devonian Series/Epoch (Eifelian) Timrzit, Morocco (precise bed location unconfirmed). Photograph by Matt Heaton, FossilEra (2023), supplied to the authors and reproduced with permission. Cincinnati Museum of Natural History and Science, accession number CMC IP100892.

## Hydrodynamic drag on tines

Gishlick and Fortey (2023) proposed that tridents may have functioned in intraspecific combat, serving to push, poke, or flip rivals, using horned rhinoceros beetles as an analogue. However, the density of saltwater is approximately 800 times greater than that of air (Gaylord, 2007), imposing substantially greater drag forces. Comparing battling rhinoceros beetles in air—where proportionately large horns impose no significant fitness costs (McCullough and Emlen, 2013)—to *Walliserops* engaging in hypothetical underwater combat poses significant problems due to the vastly different physical constraints of aquatic versus aerial environments.

Muscular power output scales with muscle cross-sectional area and volume (Rospars and Meyer-Vernet, 2016). Accordingly, *Walliserops* required substantially more muscular cross-sectional area to achieve the same dominance effects as similarly sized rhinoceros beetles. Especially for *W. trifurcatus* whose tridents were proportionately longer than rhinoceros beetles, it is unlikely that they possessed sufficient strength to wield their tridents as weapons in an aquatic environment, although this conclusion is somewhat less obvious for the other *Walliserops* species discussed.

While some limited pushing or poking may have been mechanically possible, *Walliserops* three-dimensional maneuverability for combat purposes was highly constrained by hydrodynamic drag due to the high ratio of trident dorsal surface area relative to total dorsal body surface area (including trident): approximately 17% of total surface area for *W. trifurcatus* and 10% for *W. lindoei* (Gishlick and Fortey, 2023, Fig. 1; SketchandCalc.com, 2023).

Combat among animals typically involves rapid movements, with victors often being individuals in superior physical condition (Emlen, 2008) or simply larger. If *Walliserops* engaged in hypothetical combat, combatants would have been equally constrained by the drag forces of their environment. The severe limitations on *Walliserops* muscular capacity to move their tridents multi-directionally represent a high fitness cost that reduces the likelihood for tridents' sexual-selection function.

Similarly, any inclination or declination of the *Walliserops* tridents would have significantly increased hydrodynamic drag as the animal moved through water. Analogous effects are seen in modern aquatic animals: for instance, the pectoral fins of skipjack tuna make up only 3% of total body surface area but account for 14% of total drag (Webb, 1975). Swordfish pectoral fins, though comprising 15% of the wetted surface area, increase drag by 51%, while sailfish pectoral fins contribute a 25% drag increase despite comprising just 6% of wetted area (Sagong et al., 2013). Given that the tridents of *W. trifurcatus* and *W. lindoei* constituted approximately 10–17% of total surface area and projected anteriorly into, across, or along flow, it is unlikely that these species possessed sufficient muscular strength to counteract the resulting drag forces in dynamic combat scenarios.

While the tridents of *Walliserops* likely imposed substantial drag, as discussed herein, similar surface-area appendages in aquatic animals—such as pectoral fins—often serve as mechanisms of lift and stability (Sagong et al., 2013). Although *Walliserops* tridents projected anteriorly rather than laterally, they may have functioned in a comparable stabilizing role, albeit with reduced and potentially irregular effectiveness due to turbulent flow between the tines.

Further supporting a lift-related function, bills in species such as swordfish have been shown to generate lift depending on their angle of attack relative to the flow (Lee et al., 2009). Sailfish, for example, have rounded bills comprising approximately 17% of total body length, while swordfish possess flat bills extending up to 44% (Sagong et al., 2013). By comparison, the tridents of *W. trifurcatus* approach 50% of total body length (including the trident), while those of *W. hammii*, *W. tridens*, and *W. lindoei* range from 27% to 30% (Gishlick and Fortey, 2023; Chatterton and Gibb, 2010). These size ratios suggest that *Walliserops* tridents fall well within the morphological range known to produce lift in extant aquatic species.

For comparison, the horns of *Trypoxylus dichotomus* beetles may reach approximately 40% of total body length (including the horn), or two-thirds of body length excluding horns (McCullough, 2014). These proportions are similar to those of *W. trifurcatus*, and longer than the tridents of the three other *Walliserops* species discussed herein. However, beetle horns impose little to no fitness cost in an aerodynamic environment (McCullough and Emlen, 2013).

Relative to the particularly long trident of *W. trifurcatus*, the shorter tridents of *W. hammii*, *W. tridens* and *W. lindoei* may have partially mitigated—but would not have eliminated—the hydrodynamic costs of moving an anterior projection through seawater by reducing exposed area and drag-induced torque. In proportional terms, these tridents were still only about 10–13 percentage points shorter than the comparison beetle horns, whereas that of *W. trifurcatus* was proportionately longer. Unlike horns or antlers moved through air, a submerged trident would encounter the much greater density of water. Assuming an approximately constant drag coefficient, drag force would increase with exposed area and with the square of speed relative to the surrounding water. The trident of *W. trifurcatus* would therefore be expected to impose the greatest energetic demands and bending or torsional loads during rapid, forceful combat movements.

This constraint does not exclude sexual selection, because costly structures can evolve when their reproductive benefits outweigh their energetic and mechanical costs. It does, however, mean that a combat hypothesis involving rapid underwater striking, lifting, or flipping should explain how sufficient force was generated and how the resulting drag and structural loads were resisted. These constraints may be more readily accommodated by a function involving relatively sustained orientation and trident–pygidium contact during following, in which the trident could remain more closely aligned with the flow, undergo fewer rapid accelerations, and receive mechanical support from the individual ahead. Such a function would not eliminate hydrodynamic drag, but it would place different and potentially lower dynamic loads on the trident than repeated rapid wielding.

The two functions need not have been mutually exclusive, and the trident may have served locomotor, stabilizing, tactile, reproductive, or combined roles during its evolutionary history. The present analysis does not establish which function arose first, although it identifies hydrodynamic and mechanical constraints that any combat interpretation must accommodate.

As discussed herein, lift forces generated by the tridents may have been irregular or inefficient due to turbulence between the tines, particularly during solitary locomotion. The relatively long tridents of *W. trifurcatus* and *W. lindoei*, combined with their large surface areas, may have posed disadvantages in other common scenarios. For instance, rapid accelerations caused by predator evasion or internal wave disturbances could have exposed these trilobites to destabilizing pitch or roll, making tridents counter-adaptive outside of coordinated group formations.

Together these considerations are consistent with the hypothesis that tridents served a lift and stability function that was optimally realized during collective travel. By traveling in queues and resting their tridents on the posteriors of those ahead, both followers and leaders would have benefited from enhanced roll and pitch stability. This mechanically linked arrangement allowed more consistent and stable lift generation compared to solitary movement, reinforcing the adaptive advantage of group-based locomotion.

## Tine spacing

The spacing between trident tines creates a pinnate, branched architecture that likely increases turbulence and drag relative to a continuous, elliptical, leaflike structure of equivalent surface area (Albayrak et al., 2010). However, the tines of *W. trifurcatus* and *W. lindoei* exhibit a lanceolate, leaflike morphology with likely low flexural stiffness—suggesting that they could deform passively in response to water flow. This contrasts with the apparently more rigid tines of *W. hammii* and *W. tridens*, as illustrated in Figure 8A. In fluid environments, compliant leaflike structures typically undulate with flow, reducing drag through passive reconfiguration (Albayrak et al., 2010).

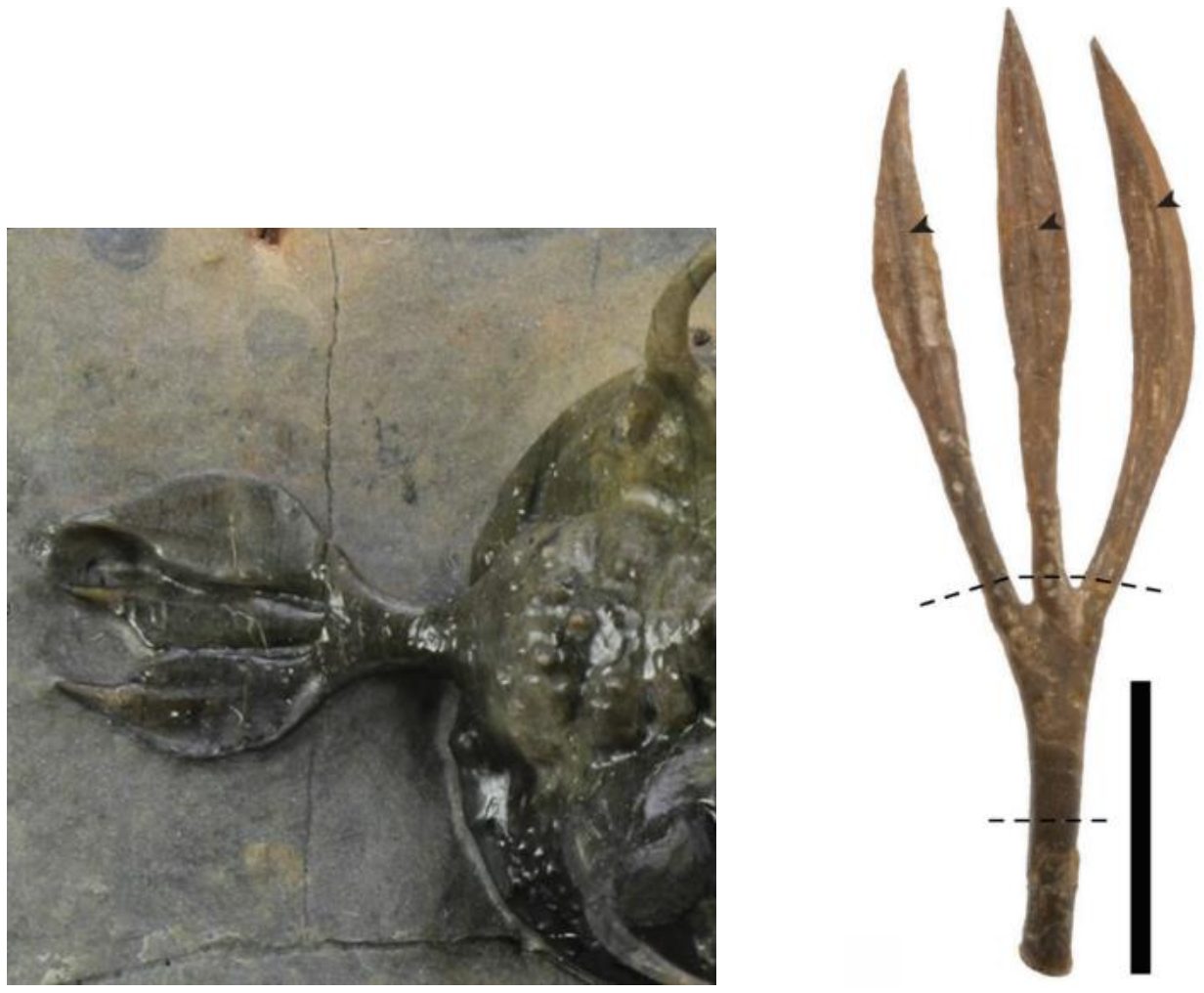

A B

**Figure 8. A.** Magnification of the trident of the second specimen from the left as shown in Figure 7. **B**. Dorsal (left) and ventral (right) views of HMNS Pl 1810; scale bar = 10 mm. Panel B adapted from Gishlick and Fortey (2023, Fig. 2A) (Creative Commons licence).

These considerations support the hypothesis that queue or cluster formations—wherein following trilobites laid their tridents dorsally on those ahead—served to mitigate destabilizing individual lift effects and reduced the risk of stall. By physically linking individuals, such formations conferred collective stability during forward motion. The reduced rigidity and leaflike morphology of the tines may have facilitated a more hydrodynamic fit between the follower's trident and the leader's pygidium, smoothing fluid flow across the connected bodies. While trilobites likely disconnected during activities such as foraging or spawning, they would have remained in close proximity to enable rapid reformation of coordinated queues for efficient collective movement.

It must be noted that trident rigidity also depends partly on the timing of exoskeletal moulting. As *Walliserops* lost entire exoskeletons during collective synchronized moulting processes (Corrales-García et al., 2020) including their tridents, the low-rigidity trident soft tissue (Drage et al., 2019) may have been temporarily ineffective for resting on the posteriors of leading trilobites during forward motion. However, *Walliserops* aggregations may have been non-migratory with spawning occurring following mass moulting (Speyer and Brett, 1985). Therefore, the presence of soft-tissue tridents during non-migratory moulting periods does not militate against tridents being used as contact structures and following mechanisms during locomotion and migration during inter-moult phases. Further modeling by computational fluid dynamics or other means may investigate these possibilities.

## Trident strain-bearing structure

The structural morphology of the trident further highlights the high biomechanical costs that would be associated with a sexually selected combat function. In *W. trifurcatus* and *W. lindoei*, the base—or haft—of the trident has a comparatively small cross-sectional area relative to the robust horn bases observed in rhinoceros beetles (Gishlick and Fortey, 2023, Fig. 4).

The tridents of *W. trifurcatus* and *W. lindoei* display a circular cross-section with relatively uniform diameter from base to the point of tine bifurcation (Gishlick and Fortey, 2023), whereas rhinoceros beetle horns are typically tapered, with wide, reinforced bases designed to withstand substantial combat stress (McCullough et al., 2015). Such tapering is an adaptive feature for resisting bending and torsional stress during combat—features largely absent in *Walliserops* tridents.

Gishlick and Fortey (2023) proposed that the trident in *W. trifurcatus* functioned as a lever, enabling flipping actions in combat. However, given the extreme length and slenderness of the trident, effective leverage would have required a stable point of support (a fulcrum), such as the substrate or the opponent's body, against which force could be applied to generate torque. Such support may have been difficult to obtain consistently during benthic combat.

Otherwise, a fulcrum was located at the cephalon-shaft connection point. According to classical mechanics, mechanical advantage is inversely proportional to the distances of the force points to the fulcrum (Balke, 2023). In *W. trifurcatus*, a fulcrum positioned near the cephalon–shaft junction would lie too far posterior to offer mechanical advantage for forward-directed flipping actions. Similarly, trident deformation was likely substantial as considerable relative forces would have been required to flip combatants in the high-drag environment. The shorter tridents of *W. tridens* and *W. hammii* may offer slightly more favorable leverage, given their more anterior pivot points and reduced moment arms. Future work may more precisely model the relative mechanical advantages conferred by the various *Walliserops* species.

## Trident-pygidium morphology

The morphology of *Walliserops* tridents aligns strikingly with the planar contours of conspecifics' pygidia, as shown in Figures 5 and 6. Slightly upward-angled tridents would have naturally conformed to the downward slope of the leading trilobite's pygidium, as discussed herein. Some specimens of *W. trifurcatus* exhibit a raised central tine, as noted by Gishlick and Fortey (2023, Supplementary Information) which may have been functionally adapted to accommodate the elevated axial lobe of a leader. However, this feature is not universal, and considerable phenotypic variation exists in the expression of the trident's three tines (Gishlick and Fortey, Supplementary Information, 2023).

As noted, the tines of *W. lindoei* and *W. trifurcatus* (but not *W. hammii* and *W. tridens*) exhibit morphological structures similar to the midrib on lanceolate leaves that divides a leaf down its centre (TreeGuideUK, 2023), as indicated in Figure 8B. Gishlick and Fortey (2023) describe these features as keels or midribs, potentially enhancing tensile strength. In mechanical engineering design, these longitudinal structures are termed "stringers" and provide strength against buckling (Spagnoli and Chryssanthopoulos, 1999).

We propose that the midrib structures, along with the leaflike curvature of the tines, functioned to accommodate the axial lobes of leaders. The tines' laterally curved structure served to accommodate the oval-shaped pygidial contour of leading trilobites. The curvature of the tines may also have allowed them to adjust passively to the contours of pygidia during lateral movements or turns by the leading trilobite, as modeled in Figure 5.

At the same time, tine midribs may have functioned as axes for anhedral (downward) leaf angles that facilitated lateral and directional maneuverability analogous to anhedral fin configurations in fish, which enhance roll responsiveness at the expense of static stability (Webb, 2005), albeit at the loss of stability exhibited by dihedral (upward) angles (Kanneganti et al., 2022). This supports the following-function hypothesis, wherein directional stability during collective travel was achieved through trident–pygidium contact. Correspondingly, anhedral tine angles allowed better maneuverability when individuals moved around in non-queue aggregations during relatively static foraging or resting periods.

*Walliserops* tine morphology should not be expected to conform precisely to the shape of the pygidia of individuals ahead. The observed variation in trident form likely reflects phenotypic plasticity—a flexible developmental response to continuously changing contact orientations that occurred during movement (Whitman and Agrawal, 2009). As followers walked and rested their tridents on the posteriors of leaders, slight shifts in alignment would have been constant. In this context, exact morphological matching would not be advantageous; rather, the observed variability is consistent with an adaptive response to a dynamic target, where flexibility and tolerance to positional variation were likely more beneficial than precise conformity.

Additionally, in the rightmost trilobite configuration shown in each of the panels in Figure 5A–D, the leader's axial lobe roughly fits between the follower's lateral tines. As the leader shifted position, its axial lobe may have rotated beneath two lateral tines, the middle straight one, and a curved exterior tine. This dynamic slotting could have allowed followers to detect subtle lateral rotations of the leader, aiding in coordinated directional changes. Such an arrangement may also have enabled the follower to maintain an unobstructed field of view on one side, depending on the orientation—paralleling behaviors observed in fish, which often position themselves at lateral angles relative to conspecifics for hydrodynamic efficiency or enhanced sensory perception (Marras et al., 2015).

While further modeling is required to demonstrate the extent to which tridents correspond to the three-dimensional components of conspecifics' pygidia, this morphological congruence between tridents and pygidia is consistent with, but not demonstrative of, evolved functional integration.

## Energetically costly variations in speed

In collective formations, trilobites would have experienced energetically costly accelerations and decelerations as they adjusted their speed to maintain position relative to others. If collisions occurred, rapid decelerations were in turn followed by accelerations. Collisions among individual *W. trifurcatus* in drafting positions were likely exacerbated on downslopes, where gravitational acceleration and reduced in-wake pressure increased speed differentials.

Several studies on fish propulsion indicate that fluctuating speeds are more energetically demanding than swimming steadily at a constant average speed (Webb, 1991; Boisclair and Tang, 1993; Krohn and Boisclair, 1994). Although intermittent propulsion typically requires more energy than constant-speed movement, it may offer energy advantages when gliding phases are incorporated, with gliding shown to be more energetically beneficial in flying and swimming animals than in those that move terrestrially (Kramer and McLaughlin, 2001).

Therefore, *Walliserops* may have formed seafloor queues with tridents in contact not only to enhance collective locomotor stability, but also to minimize the energy costs of collisions and variable speeds while also exploiting the energy-saving benefit of drafting.

It is reasonable to infer that trilobites could respond behaviourally to changing locomotor demands, with followers adjusting their positions to exploit lower-drag wake regions. Such positioning may have allowed individuals with lower sustainable power output to maintain the group's pace and thereby promoted collective cohesion. If the pace nevertheless exceeded a follower's physiological capacity—even with the benefit of drafting—that individual could have fallen out of the group or joined a slower subgroup. Repeated differences of this kind could have sorted groups according to age or locomotor capacity and imposed selection on sustained locomotor performance (Trenchard et al., 2017).

At sufficiently high relative flow speeds, the energetic advantage of occupying a leader's wake could have favoured close-following behaviour before specialized tridents evolved. Because drag reduction was greatest at short following distances, this behaviour would also have exposed followers to collisions, repeated speed adjustments and positional instability. A simple anterior projection that reduced direct collisions, regulated fore–aft spacing or helped followers remain within drag-reduced wake positions could therefore have conferred an individual advantage. Progressive elaboration into a three-tined structure may have provided additional mechanical stabilization: the central and lateral tines may have engaged different regions of the pygidium or axial lobe of the individual ahead, limiting lateral slippage and relative pitch, roll and yaw.

The precise substrate conditions encountered by *W. trifurcatus* cannot be reconstructed, although the broader Anti-Atlas encompassed varied shelf and platform environments, as discussed above (Bault et al., 2022; Wendt, 2021). Where local seafloor relief or rocky or biogenic obstacles were present, improved coupling and stabilization could also have reduced repeated decelerations, directional corrections and re-accelerations. The trident may therefore have functioned not merely as a spacer but as a multidirectional stabilizing structure that helped followers retain energetically favourable wake positions. Comparative mechanical and hydrodynamic modelling of tridents and simpler anterior projections will be required to test this proposed evolutionary pathway.

## Migrating *Walliserops*

Locomotion is among animals' most energetically costly activities (Brown et al., 2023). As discussed herein, collective locomotion strategies that reduce energy expenditure confer clear evolutionary advantages (see also Mirzaeinia and Hassanalian, 2019).

As discussed, due to their spiny exoskeletons and limited hydrodynamic streamlining, *Walliserops* are believed to have been benthic organisms, walking along the seafloor (Van Viersen and Kloc, 2022). However, the extent of migratory behavior in *Walliserops* remains uncertain, as direct evidence is lacking.

Nonetheless, trilobite migration has been inferred in other taxa based on fossilized linear formations—suggesting coordinated travel to spawning grounds during reproductive periods (Błażejowski et al., 2016). If *Walliserops* engaged in migratory movement, they likely traveled in close-knit formations, exploiting drafting to reduce energy costs. As modeled in Equations (13) through (15), isolated individuals would have struggled to maintain the pace of those benefiting from collective energy savings—rendering isolation selectively disadvantageous. The configuration of four *W. lindoei* shown in Figure 7 offers a suggestion of this linear grouping, although it is far from conclusive and better evidence is needed.

## Conclusion

The substantial hydrodynamic costs associated with moving a trident through seawater raise difficulties for interpreting the structure primarily as an effective combat weapon. We instead propose an alternative hypothesis: that the trident acquired adaptive value, at least initially, by facilitating energy-saving drafting and positional stabilization during group locomotion, whether during long-distance migration or shorter-distance travel.

The model suggests that elevated trident–body configurations could have reduced modeled speed and potentially impaired the stability of solitary individuals because the accompanying increase in drag outweighed the reduction in apparent-weight-related power demand. Lift could reduce apparent-weight-related loading in both solitary and grouped individuals, but its effects may have been more effectively exploited in queues linked by trident–pygidium contact. Such contact could have limited lift-induced pitch and roll instability, while drafting reduced followers' drag-related power requirements and helped them maintain the group's pace. The trident's proposed collective advantage may therefore have arisen from the complementary effects of lift, drafting and mechanical stabilization rather than from lift alone. The model does not, however, establish that these combined benefits necessarily exceeded the additional drag cost of the trident.

The $Fr \approx 1$ condition at a modeled relative flow speed of approximately 20.63 cm s$^{-1}$ provides an operational benchmark—not a sharp physical threshold—for a regime in which hydrodynamic drag and the potential energetic benefits of drafting became increasingly important. In still water, drag accounts for an increasing share of modeled power demand above this benchmark and toward the assumed high-output speed of 40 cm s$^{-1}$, making reduced-drag wake positions progressively more advantageous. Drafting did not, however, become important at one fixed locomotion speed: opposing currents or elevated angles of attack could increase drag and make drafting savings consequential at otherwise moderate substrate-relative speeds, whereas following currents could reduce them. Effects of substrate slope were not modeled.

Aside from the suggestive specimen shown in Figure 7, no fossil evidence directly demonstrates queue formation in *Walliserops*. Nevertheless, the linear formations of *Ampyx*, involving contact through cephalic projections, establish that comparable queueing arrangements occurred among trilobites (Vannier et al., 2019). Although this precedent does not demonstrate the behaviour in *Walliserops*, its combination with the proposed mechanical compatibility of trident–pygidium contact and the modeled hydrodynamic benefits supports trident-assisted queueing as a testable working hypothesis.

*Walliserops* may therefore have evolved from ancestors that lacked tridents but already exploited drafting, an energy-saving behaviour widespread in nature (Trenchard and Perc, 2016). Close following would also have increased the risk of collisions, repeated speed adjustments and positional instability. Variants bearing anterior projections that improved spacing and alignment or helped followers remain within reduced-drag wakes could consequently have gained an advantage. Progressive elaboration of these projections may have produced the tridents of *W. trifurcatus* and *W. lindoei*. The trident did not create drafting; rather, it may have enabled followers to exploit that pre-existing benefit more effectively by stabilizing trident–pygidium contact, limiting pitch and roll, reducing positional corrections and permitting more stable use of lift. These effects could have helped followers maintain the group's pace at reduced power while increasing queue cohesion during both short- and long-distance movement.

From a broader evolutionary perspective, the present model identifies no clear net advantage of the trident for isolated locomotion. Its principal proposed benefits instead emerge in a group context through drafting, coordination and mechanical stabilization. This group-dependent interpretation contrasts with the primarily combat-based hypothesis of Gishlick and Fortey (2023), but does not exclude the possibility that the trident served multiple functions or was later exapted for combat or sexual selection. Morphological features promoting cooperative behaviour among conspecifics remain less studied than interspecific mutualistic or symbiotic adaptations (e.g., Leigh, 2010), and the proposed sequence remains to be tested.

Future research could test this hypothesis using three-dimensional models of *W. trifurcatus* in water flumes and computational fluid dynamics simulations. Such work could quantify drag, lift, wake structure and stability in solitary and linked configurations. Similar modelling of *Ampyx priscus*, for which fossil evidence of linear formations exists, could evaluate the effects of queueing and cephalic projections. Comparative studies of *W. hammii*, *W. tridens* and *W. lindoei* could assess functional variation among trident morphologies, while analyses of other queue-forming or anteriorly projected trilobites, including *Psychopyge elegans* and *Quadrops flexuosa*, could test the broader applicability of the hypothesis.

## Author Contributions

HT wrote the paper and conducted research. CB commented on the paper, provided research details, nomenclature and certain specific paragraphs. MP reviewed and provided comments on the paper.

M.P. was supported by the Slovenian Research and Innovation Agency (Javna agencija za znanstvenoraziskovalno in inovacijsko dejavnost Republike Slovenije) (Grant No. P1-0403).

## Acknowledgements

We thank the three reviewers for their thorough and constructive comments, which helped improve the substance and clarity of the manuscript. We also thank Matt Heaton and Shawn Rodgers, from whom we acquired the *Walliserop*s specimen shown in Fig. 7, and Brenda Hunda of the Cincinnati Museum Center for curatorial assistance. C.B. gratefully acknowledges support from the National Geographic Society, which funded broader research on Moroccan trilobite taphonomy; and Błażej Błażejowski for helpful discussions concerning trilobite behaviour. Finally, we acknowledge the late Adrian Kin and Andrzej Radwanski, whose pioneering observations first drew scientific attention to the phenomenon of trilobite queuing.

## Declaration of AI and AI-assisted technologies in the writing process

During the preparation of this work, the authors used ChatGPT (OpenAI) to assist with language editing and readability and with the preparation of Figures 2–4. After using this tool/service, the authors reviewed and edited the content as needed and take full responsibility for the content of the publication.